\documentclass[a4paper]{jpconf}
\usepackage{graphicx}

\usepackage[reqno,sumlimits,intlimits]{amsmath}
\usepackage{amstext,amsthm,amssymb,amsfonts}
 
\usepackage[english]{babel}
\usepackage[english]{cleveref}
\usepackage{overpic}
\usepackage{paralist}

\newcommand{\AuAu}{Au\,+\,Au}
\newcommand{\PbPb}{Pb\,+\,Pb}
\newcommand{\ProtonProton}{p\,+\,p}
\newcommand{\BAMPS}{\textsc{Bamps}}

\newcommand{\GeVperfemtometrecubic}{\,\textrm{GeV}/\textrm{fm}^3}
\newcommand{\fm}{\,\textrm{fm}}
\newcommand{\TeV}{\,\textrm{TeV}}
\newcommand{\GeV}{\,\textrm{GeV}}
\newcommand{\AGeV}{ $\textrm{A}\,\textrm{GeV}$}
\newcommand{\ATeV}{ $\textrm{A}\,\textrm{TeV}$}

\begin{document}
\title{RHIC and LHC phenomena with an unified parton transport}

\author{Ioannis Bouras$^a$, Andrej El$^a$, Oliver Fochler$^a$, Felix Reining$^a$,
Florian Senzel$^a$, Jan Uphoff$^a$, Christian Wesp$^a$, Zhe Xu$^b$ and Carsten Greiner$^a$}

\address{$^a$Institut f\"ur Theoretische Physik, Johann Wolfgang Goethe-Universit\"at\\
Max-von-Laue-Str.\ 1, D-60438 Frankfurt am Main, Germany}

\address{$^b$Department of Physics, Tsinghua University \\
Beijing 100084, China}

\ead{carsten.greiner@th.physik.uni-frankfurt.de}

\begin{abstract}
We discuss recent applications of the partonic pQCD based cascade model BAMPS
with focus on heavy-ion phenomeneology in hard and soft momentum range.
The nuclear modification factor as well as elliptic flow are calculated in
BAMPS for RHIC end LHC energies. These observables are also discussed
within the same framework for charm and bottom quarks. Contributing to the
recent jet-quenching investigations we present first preliminary
results on application of jet reconstruction algorithms in BAMPS.
Finally, collective effects induced by jets are investigated: we
demonstrate the development of Mach cones in ideal matter as well
in the highly viscous regime.
\end{abstract}

\section{Introduction}

In collisions of heavy ions at ultrarelativistic energies at the Relativistic
Heavy-Ion Collider (RHIC) and the Large Hadron Collider (LHC) a new state of
matter, the quark-gluon plasma (QGP), has been created. Although the QGP
is not available for direct observation, its properties can be deduced from the
measurement of the produced hadrons in the final state.

The large values of the measured hadronic elliptic flow $v_2$ \cite{Adler:2003kt,Adams:2003am,Back:2004mh},
which is the second coefficient of the
Fourier series of the azimuthal particle multiplicity, suggests that equilibration of
quarks and gluons occurs on a very short time scale $\le 1$ fm/c. This also suggests
that the shear viscosity over entropy density ratio $\eta/s$ of the QGP is very small,
which means that the QGP behaves like a nearly perfect fluid. All these conclusions can be
drawn from comparison of experimental results with hydrodynamic calculations.
However, an understanding of the mechanism of fast thermalization can not be achieved in the scope
of hydrodynamic models. The early pre-equilibrium dynamics of the QGP must be
studied in the scope of the kinetic theory.

In contrast to the hydrodynamic approach, kinetic transport theory is a microscopic theory
and thus allows to study processes of soft and hard processes simultaneously. This
is in particular important for detailed understanding of further properties
of the quark-gluonic medium, such as the suppression of jets and heavy-quarks. Suppression
of jets, also known as jet quenching, is quantified by comparing the hadron
multiplicities measured in heavy-ion collisions with appropriately scaled multiplicities
from $p+p$ collisions \cite{Adams:2003kv,Adler:2002xw,Adcox:2001jp}. In addition, very
exciting jet-associated particle correlations were observed \cite{Wang:2004kfa},
which might be the result of a conical emission off propagating shock waves in
form of Mach Cones. These Mach Cones might be induced by high-energy partons
traversing the expanding medium \cite{Stoecker:2004qu}. Observations
of these effects is consistent with the picture of a nearly perfect fluidity of the QGP.

The  kinetic transport model BAMPS (Boltzmann Approach to Multiparton Scatterings)
\cite{Xu:2004mz} has been developed to provide a unified description of dynamics of
the early QGP stage of heavy-ion collisions (HIC) including perturbative QCD based
elastic and inelastic processes. BAMPS has been applied to provide explanation
of fast thermalization on a very short time scale $\le 1$ fm/c \cite{El:2007vg}
as well as a small value of $\eta/s \approx 0.08 - 0.2$ for $\alpha_s = 0.6 - 0.3$
\cite{Xu:2007ns,El:2008yy}. Furthermore recent calculations with BAMPS
provide results on elliptic flow \cite{Xu:2007jv,Xu:2008av} and jet
quenching \cite{Fochler:2008ts} at RHIC energies, which is for the first time
done in a consistent and fully pQCD--based microscopic transport model. In addition,
BAMPS has been used in certain works as a reference for hydrodynamic calculations.
This opens the possibility to study hydrodynamic phenomena for arbitrary viscosity.

In these proceedings we discuss application of BAMPS to describe a number
of phenomena observed in the recent heavy-ion experiments. In Sec.~\ref{section_RAA}
we introduce calculations of the nuclear modification factor $R_{AA}(p_T)$
for RHIC and LHC conditions. In Sec.~\ref{section_heavyQuarks} BAMPS
results on elliptic flow and suppression of charm and bottom quarks for RHIC
and LHC energies are introduced. In Sec.~\ref{section_jetReconstruction}
for the first time in the framework of BAMPS we introduce preliminary results on
application of jet reconstruction algorithms. Finally in Sec.~\ref{section_machCones}
the formation and propagation of shock waves in form of Mach Cones
are discussed for a wide range of viscosity to entropy density ratio
$\eta/s$.

\section{Nuclear modification factor and elliptic flow from partonic transport simulations}
\label{section_RAA}

As established in \cite{Xu:2007jv,Xu:2008av,Fochler:2008ts} the partonic medium in BAMPS
simulations of ultra relativistic heavy ion collisions features a small ratio of the shear
viscosity to the entropy, $\eta/s$, and develops a strong collectivity with an integrated
$v_{2}$ that is in good agreement with experimental results over a large centrality range
for a fixed coupling of $\alpha_{s}=0.3$ and a kinetic freeze-out energy density
$\varepsilon_{c} = 0.6 \GeVperfemtometrecubic{}$. These parameters are
used for all calculations that are presented in this section.

\begin{figure}[tbh]
 \centering
 \includegraphics[width=0.7\textwidth]{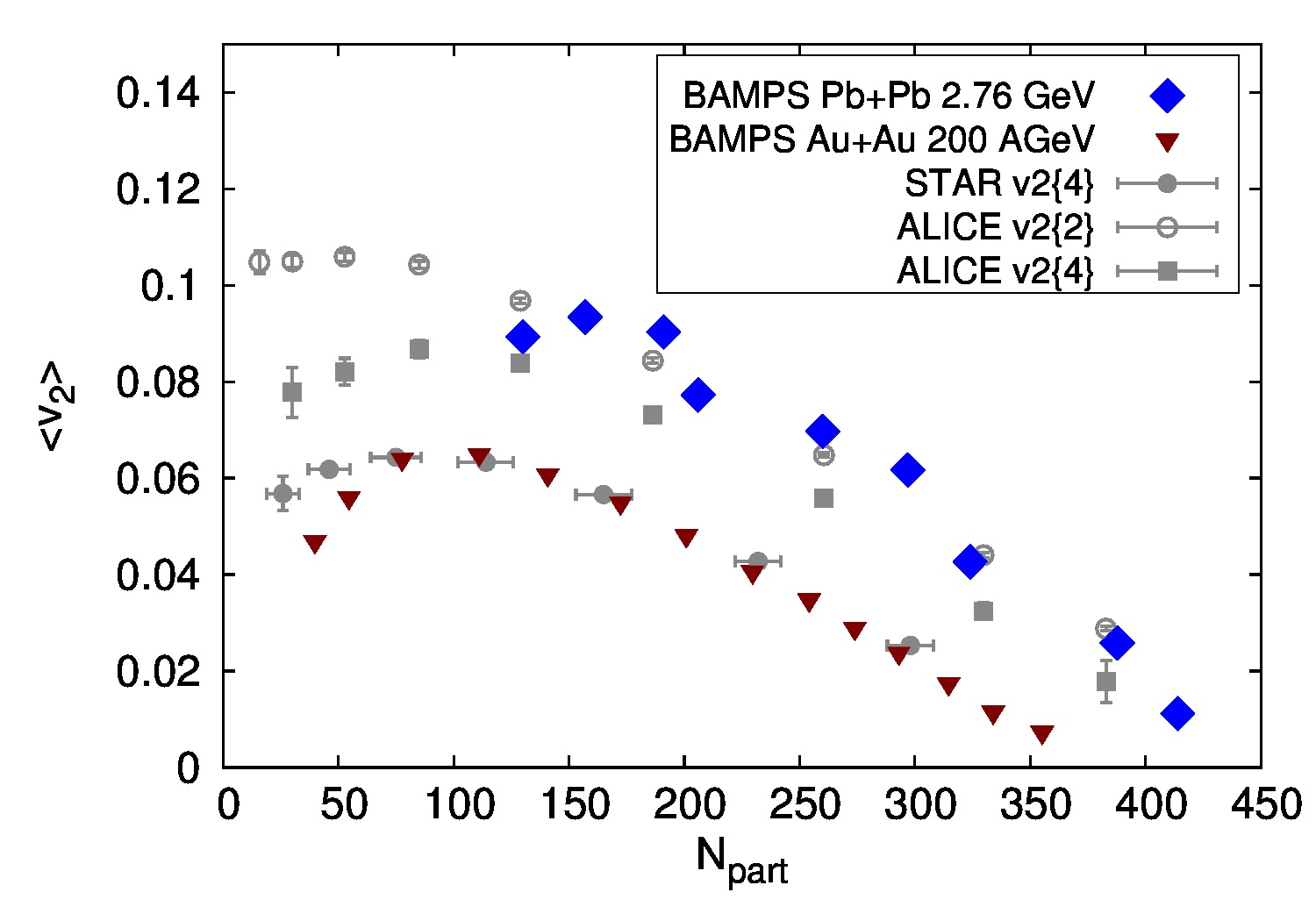}
 \caption[]{Integrated partonic $v_{2}$ from BAMPS as a function of $N_{\text{part}}$
for \PbPb{} at 2.76 \ATeV  ($|y|<0.8$) and \AuAu{} at 200 \AGeV  ($|y|<0.5$)
compared to the measured $v_{2}$ of charged particles from ALICE \cite{Aamodt:2010pa} and from STAR \cite{Adams:2004bi}.}
\label{fig:v2_LHC_RHIC}
\end{figure}

Fig.\ \ref{fig:v2_LHC_RHIC} shows the integrated partonic $v_{2}$ as a function of centrality
from simulations of \AuAu{} at 2.76 \ATeV and of \AuAu{} at 200 \AGeV compared to
experimental data from the ALICE experiment \cite{Aamodt:2010pa} at LHC and from the STAR
experiment \cite{Adams:2004bi} at RHIC. Using the same set of parameters
($\alpha_{s}=0.3$, $\varepsilon_{c} = 0.6 \GeVperfemtometrecubic$) that have been
fixed to the RHIC data \cite{Xu:2007jv,Xu:2008av} also the integrated elliptic flow at LHC can
be described over a large range in centrality. Accordingly the simulated differential $v_{2}$
of \PbPb{} collisions at LHC energies shows no significant deviation from the \AuAu{} results
at 200 \AGeV in the low to intermediate $p_{T}$ region which is also in agreement with
experimental findings \cite{Aamodt:2010jd}.

\begin{figure}[tbh]
 \centering
 \includegraphics[width=0.7\textwidth]{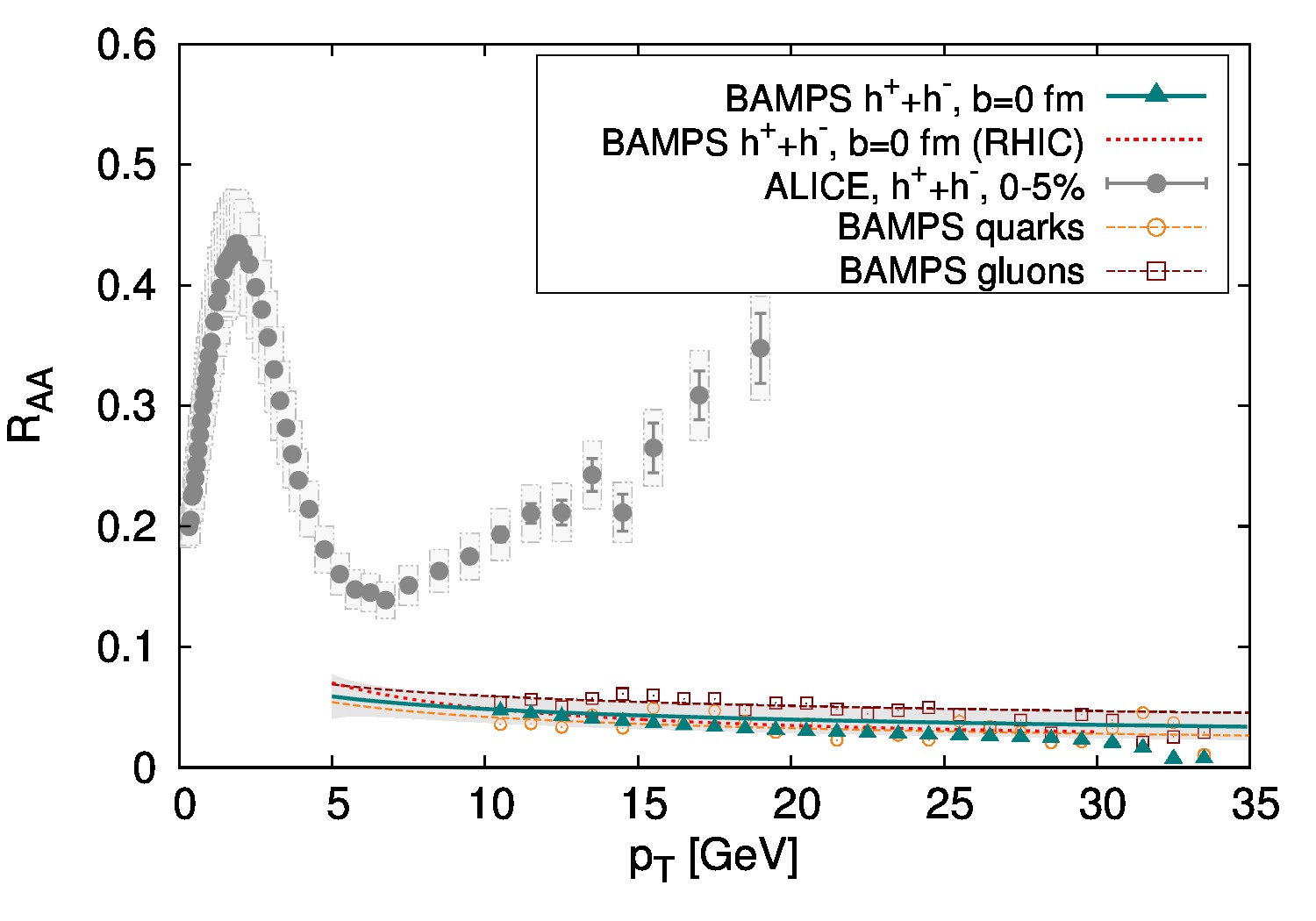}
 \caption[]{Nuclear modification factor $R_{AA}$ of charged hadrons, gluons and quarks from
BAMPS simulations of \PbPb{} at $b=0\fm{}$ compared to results from ALICE for
0-5\% central \PbPb{} collisions \cite{Aamodt:2010jd}. Lines indicate
$R_{AA}$ computed from fits to the simulated parton spectra, while symbols indicate $R_{AA}$
computed directly from the parton spectra as obtained from BAMPS. For comparison the $R_{AA}$
of charged hadrons from simulations of \AuAu{} at 200 \AGeV and $b=0 \fm{} $ is also shown.}
\label{fig:RAA_LHC}
\end{figure}

One of the main virtues of the transport model BAMPS is that it allows for the investigation
of different observables within a consistent framework and consequently also the nuclear
modification factor, $R_{AA}$, is studied using the same parameters that provide a description
of the elliptic flow ($\alpha_{s}=0.3$, $\varepsilon_{c} = 0.6 \GeVperfemtometrecubic$).
Fig.\ \ref{fig:RAA_LHC} shows the nuclear modification factor $R_{AA}$ obtained from BAMPS
simulations of central, 0-5\%, \PbPb{} collisions at
2.76 \ATeV. The results are both shown on the partonic level for gluons and light quarks
and on the hadronic level for neutral pions based on AKK fragmentation functions
\cite{Albino:2008fy}. The suppression of high-$p_{T}$ particles in simulations with BAMPS
is distinctly stronger than the experimentally observed suppression, which is also observed
in simulations of $R_{AA}$ at RHIC energies \cite{Fochler:2010wn}. Additionally the rise
towards larger transverse momenta that is present in the LHC data is not reproduced.

The strong quenching observed in BAMPS calculations is due to the energy loss in
$2 \rightarrow 3$ interactions that include an effective implementation of the LPM
effect \cite{Landau:1953um} via a mean free path-dependent cutoff \cite{Xu:2004mz,Fochler:2008ts}.
The strong quenching is then caused by \cite{Fochler:2010wn}
\begin{inparaenum}[\itshape a\upshape)]
\item a strong energy loss that is caused by a complex interplay of the Gunion-Bertsch
matrix element and the effective implementation of the LPM effect \cite{Fochler:2010wn},
\item a conversion of quark into gluon jets in $2 \rightarrow 3$ interactions and
\item a small difference in the energy loss of gluons and quarks caused by the iterative
computation of interaction rates required by the inclusion of the LPM cutoff.
\end{inparaenum}

Thus, while the collectivity of the medium can be well described within the current approach,
the quenching of high-$p_{T}$ particles is overestimated. Future studies will therefore focus
on the implementation of a running coupling for light quarks and gluons and also systematically
explore the modeling of the LPM effect. These modifications are qualitatively expected to bring
the results for the nuclear modification factors into better agreement with experimental data.

\section{Elliptic flow and suppression of heavy quarks}
\label{section_heavyQuarks}

\begin{figure}
\begin{minipage}[t]{0.49\textwidth}
\centering
\includegraphics[width=1.0\textwidth]{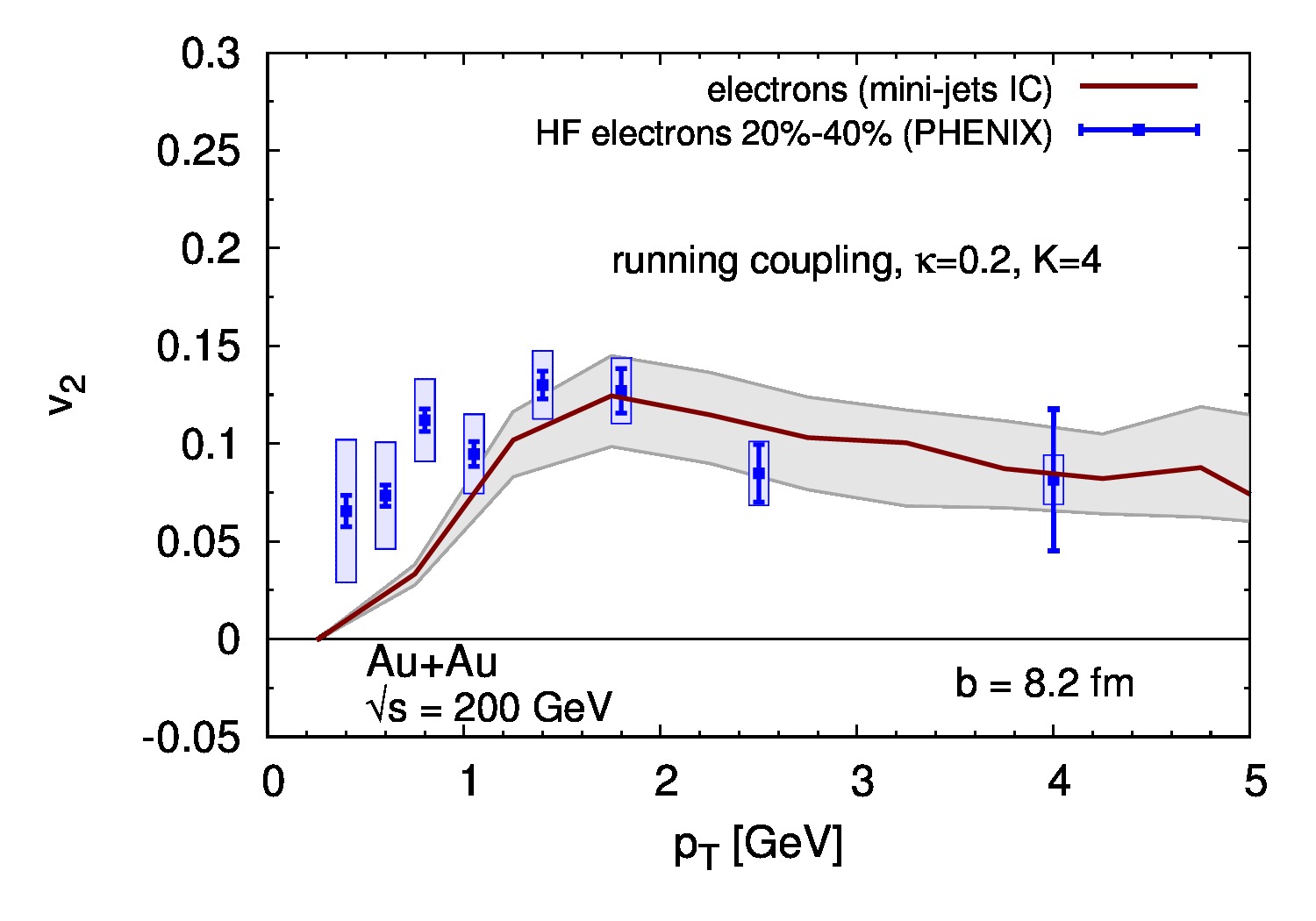}
\end{minipage}
\hfill
\begin{minipage}[t]{0.49\textwidth}
\centering
\includegraphics[width=1.0\textwidth]{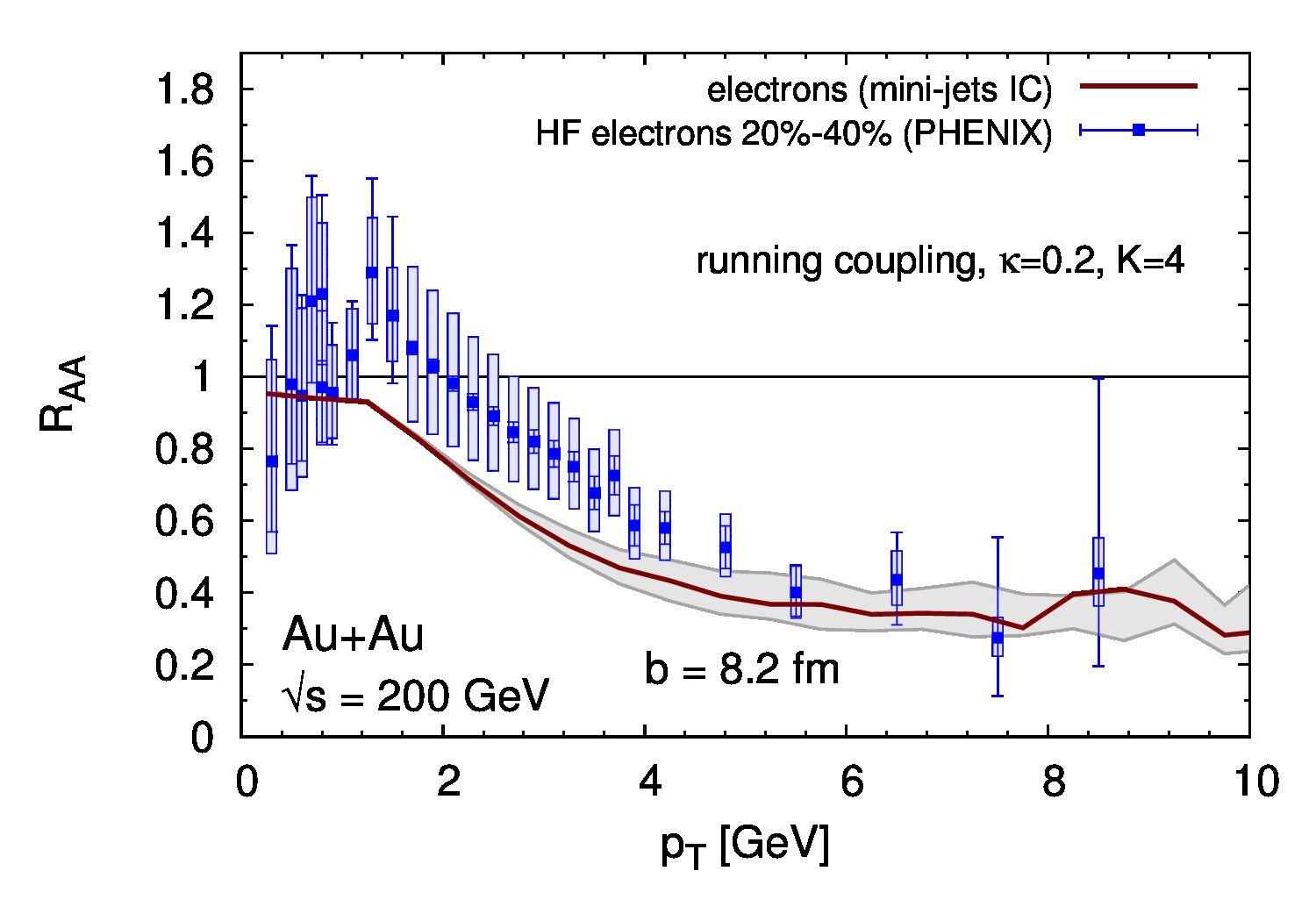}
\end{minipage}
\caption{Elliptic flow $v_2$ (left) and nuclear modification factor $R_{AA}$ (right) of
heavy flavor electrons for Au+Au collisions at RHIC with an impact parameter of $b=8.2 \, {\rm fm}$
together with data \cite{Adare:2010de}.
 The elastic cross section of $gQ \rightarrow gQ$ 
is multiplied with the factor $K=4$ to mimic the influence of radiative processes.
}
\label{fig:v2_raa_rhic}
\end{figure}

Heavy quarks are a good probe to study the properties of the QGP.
They are well calibrated in a sense that they are produced entirely
in the early stage of the heavy ion collision due to their large
mass \cite{Uphoff:2010sh} and are also tagged during hadronization
due to flavor conservation. Whereas heavy quarks at RHIC can only
be measured indirectly via heavy flavor electrons, at LHC for the
first time it is possible to reconstruct $D$ mesons and, therefore,
receive information only about charm quarks.

\begin{figure}
\begin{minipage}[t]{0.49\textwidth}
\centering
\begin{overpic}[width=1.0\textwidth]{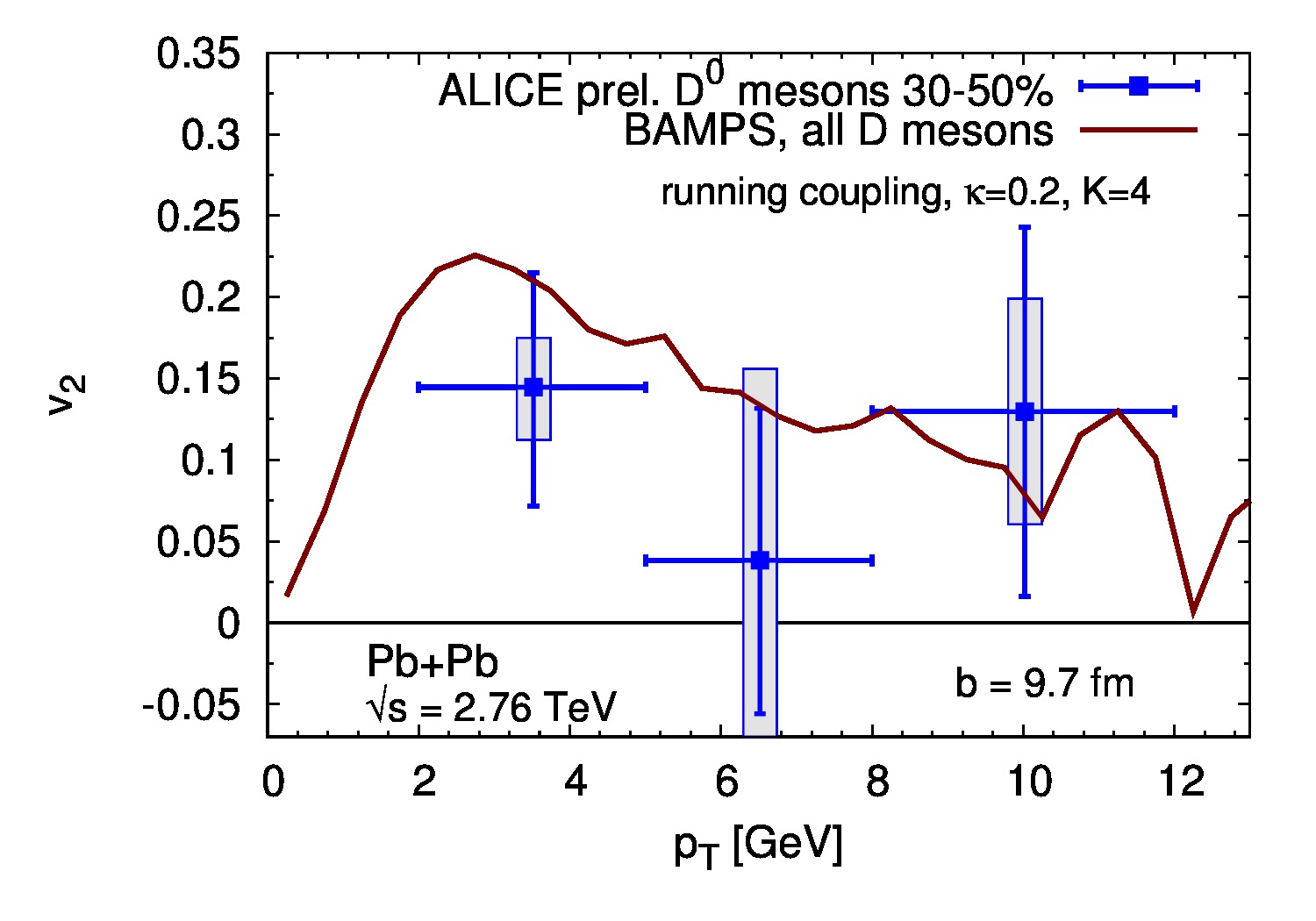}
\put(48,48){\footnotesize preliminary} 
\end{overpic}
\end{minipage}
\hfill
\begin{minipage}[t]{0.49\textwidth}
\centering
\begin{overpic}[width=1.0\textwidth]{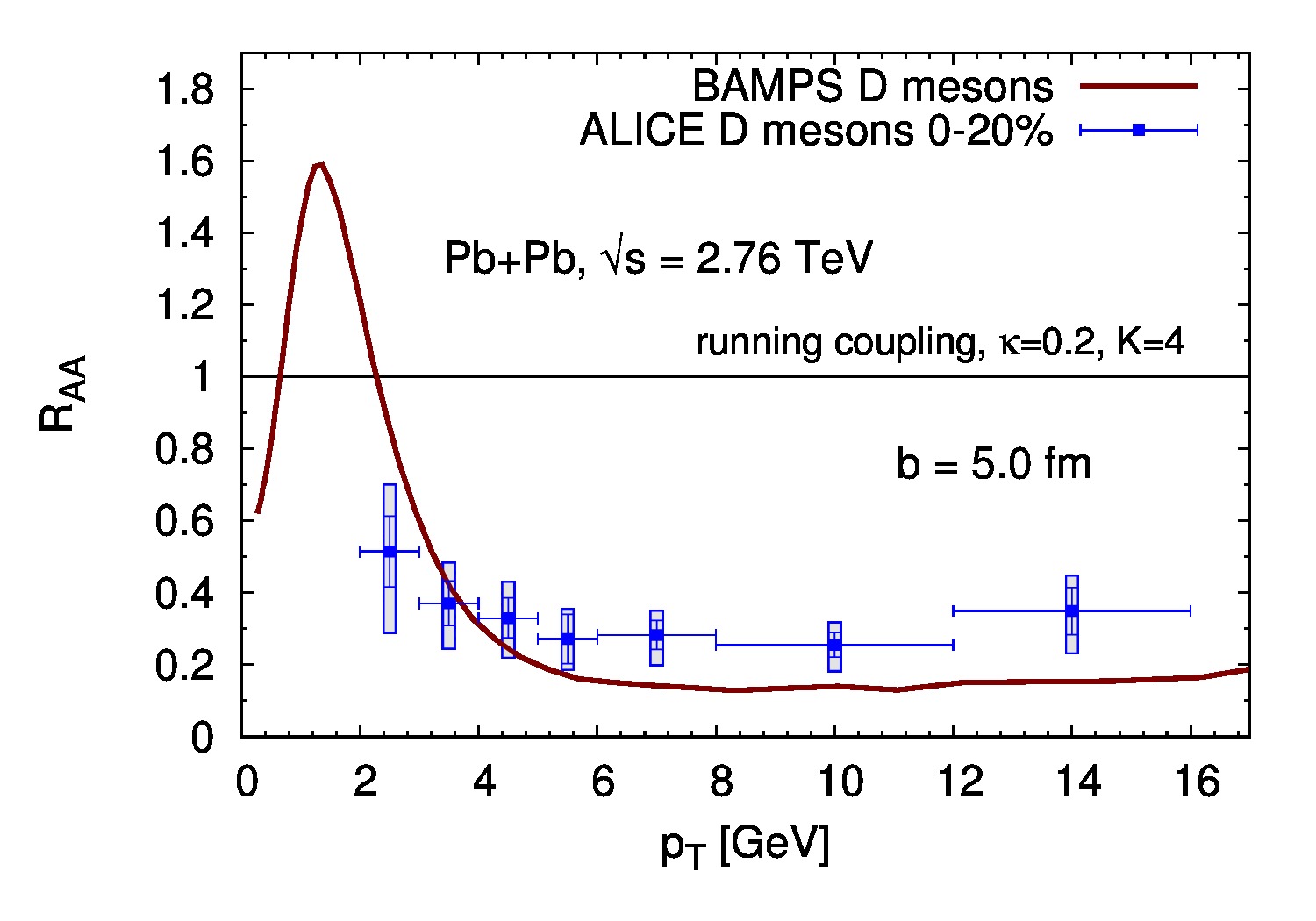}
\put(63,27){\footnotesize preliminary} 
\end{overpic}
\end{minipage}
\caption{Preliminary BAMPS results on elliptic flow $v_2$ (left) and nuclear
modification factor $R_{AA}$ (right) of $D$ mesons at Pb+Pb collisions at LHC
with an impact parameter $b$ together with data \cite{Abelev:2012nj, Bianchin:2011fa}.
The cross section of $gQ \rightarrow gQ$ is multiplied with the factor $K=4$.
}
\label{fig:v2_raa_d_meson_lhc}
\end{figure}

The heavy flavor electron data from RHIC
\cite{Abelev:2006db,Adare:2006nq,Adare:2010de} and the heavy flavor
electron, muon and $D$ meson data from LHC \cite{Masciocchi:2011fu,Abelev:2012nj}
show that the suppression of heavy quarks is on the same
order as for light quarks. From the theory perspective it
was thought that radiative processes involving heavy quarks
are suppressed due to the dead cone effect \cite{Dokshitzer:2001zm,Abir:2011jb},
which means that gluon radiation at small angles is suppressed and,
therefore, the energy loss is smaller compared to light partons.
Elliptic flow $v_2$ measurements of particles associated with open
heavy flavor also show that heavy quarks interact strongly with the
other particles of the medium. Whether these observations can be explained
by collisional or radiative energy loss or other effects is currently in
debate.

The elliptic flow $v_2$ and the nuclear modification factor $R_{AA}$ are
important observables for heavy quarks. Although those particles are rare
probes, both observables are experimentally accessible for fragmentation
and decay products of heavy quarks such as $D$ mesons or heavy flavor
electrons. The $R_{AA}$ reflects how much energy heavy quarks lose in
the QGP. The $v_2$ is large if heavy quarks interact often with the medium
and pick up its collective flow.

All the calculations for heavy quarks in this section are done with a
running coupling and an improved Debye screening. The latter means that
the screening mass of the $t$ channel of elastic scatterings is determined
such that the energy loss matches the energy loss of a heavy quark
calculated within the hard thermal loop approach. More information how
this matching is done can be found in Ref.~\cite{Gossiaux:2008jv,Peshier:2008bg,Uphoff:2011ad,Meistrenko:12}. 

Detailed studies in BAMPS \cite{Uphoff:2011ad, Uphoff:2011aa, Fochler:2011en, Uphoff:2010bv, Uphoff:2010sy}
show that elastic energy loss of heavy quarks alone is not compatible
with the experimental data at RHIC and LHC. However, elastic energy
loss explains a significant portion of the overall suppression. If we
employ a running coupling and improved Debye screening the experimental
data for both $v_2$ and $R_{AA}$ for both RHIC and LHC can be explained
if the elastic cross section is multiplied with the artificial factor
$K = 4$. This indicates that radiative energy loss should be three times
larger than the elastic energy loss. However, this must be checked in a
forthcoming study. First results on implementing radiative energy loss
of heavy quarks in BAMPS look promising \cite{Uphoff:2011aa}.

Figure~\ref{fig:v2_raa_rhic} compares our results of the $v_2$ and $R_{AA}$
at RHIC to the heavy flavor electron data from Ref.~\cite{Adare:2010de}. 
The agreement with the experimental data is very good for both observables
if one employs a factor $K=4$ for the elastic cross section to mimic the
effect of radiative energy loss.
At LHC for the first time it is possible to reconstruct $D$ mesons and,
therefore, distinguish between charm and bottom quarks. In Fig.~\ref{fig:v2_raa_d_meson_lhc}
our results on $D$ mesons is compared to data from ALICE.
For the same parameters, that describe the RHIC data, a good agreement is
also found at LHC. The suppression of $D$ mesons at LHC is slightly larger
than the data. This can be due to a different relation between collisional
and radiative processes at LHC compared to RHIC or due to the fact that
we represent the rather large centrality class $0-20$~\% by only one impact
parameter. We note that muon data from charm and bottom quarks at forward
rapidity is also well described for the same parameters \cite{Uphoff:2011aa}.

\section{Jet reconstruction within \BAMPS}
\label{section_jetReconstruction}

Another observable to determine the parton energy loss inside
a heavy-ion medium is the reconstruction of full dijets. The
initial hard scattering processes of the approaching nucleons
lead to back-to-back parton pairs, which gain a high amount of
virtuality during these scattering processes. In the subsequent
evolution of partons, they try to decrease their virtuality
by splitting processes like
$ \textrm{q} \rightarrow \textrm{q} \textrm{g} $ or
$ \textrm{g} \rightarrow \textrm{g} \textrm{g}$, which can be
described by the DGLAP evolution equation
\cite{Gribov:1972ri,Dokshitzer:1977sg,Altarelli:1977ts}. These fragmentation
processes lead to particle showers with a broad angle and momentum
distribution. In order to provide a description of the energy loss mechanism
inside the created medium, jet reconstruction
methods \cite{Salam2009,Cacciari2008,Cacciari2010} are used. They
combine single shower particles to a common ``full jet'' based
on their distance $\Delta R = \sqrt{(\Delta y)^2+(\Delta \phi)^2}$
to the jet axis.

\begin{figure}[tb]
\centering
\begin{overpic}[width=0.8\textwidth]{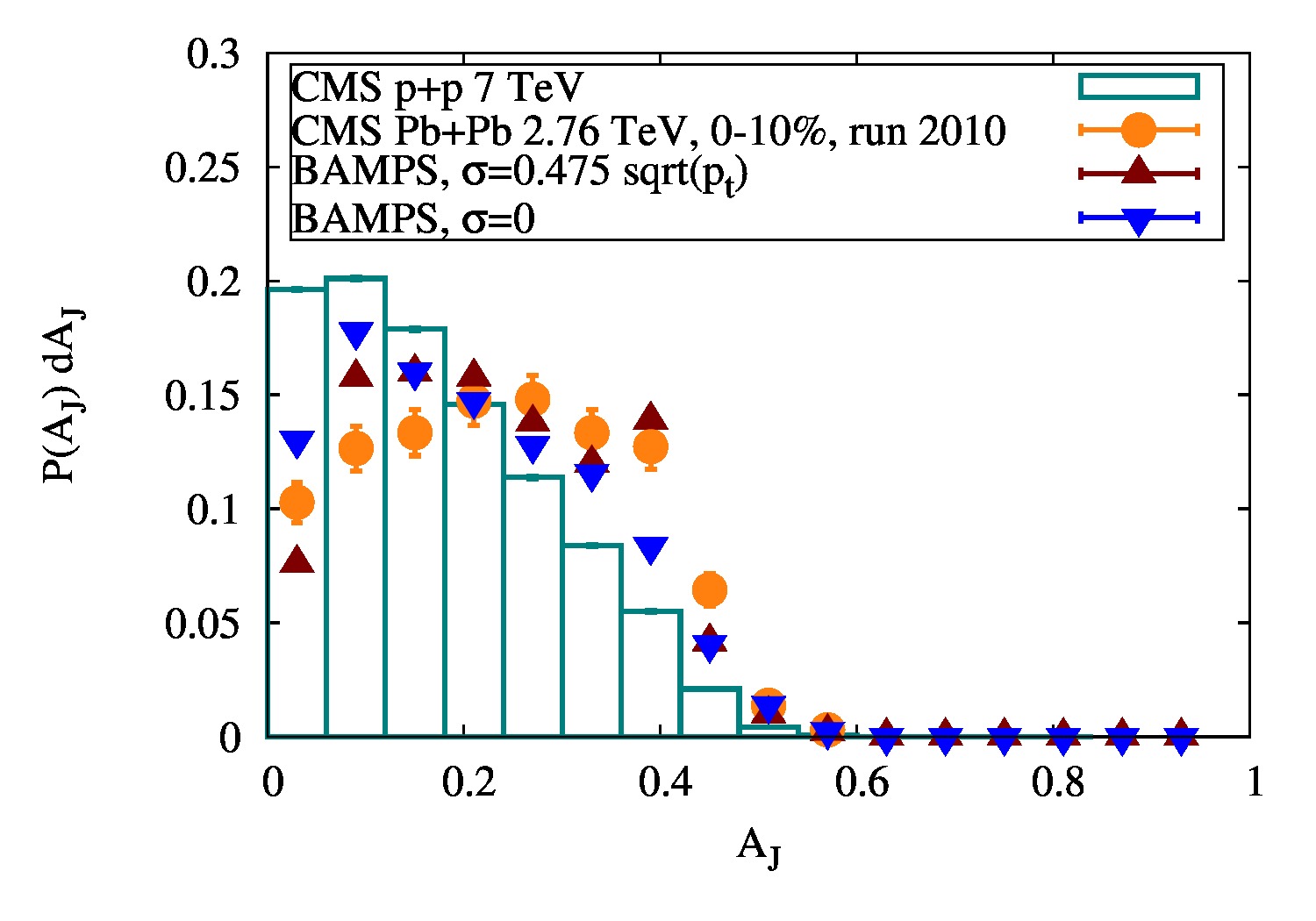}
\put(65,30){preliminary} 
\end{overpic}
\caption{Momentum imbalance $A_J$ in central \PbPb collisions at the LHC
with and without gaussian smearing for $\alpha_s = 0.3$ \cite{CMS2011}.}
\label{fig:Aj_gauss}
\end{figure}

\begin{figure}[tb]
\centering
\begin{overpic}[width=\textwidth]{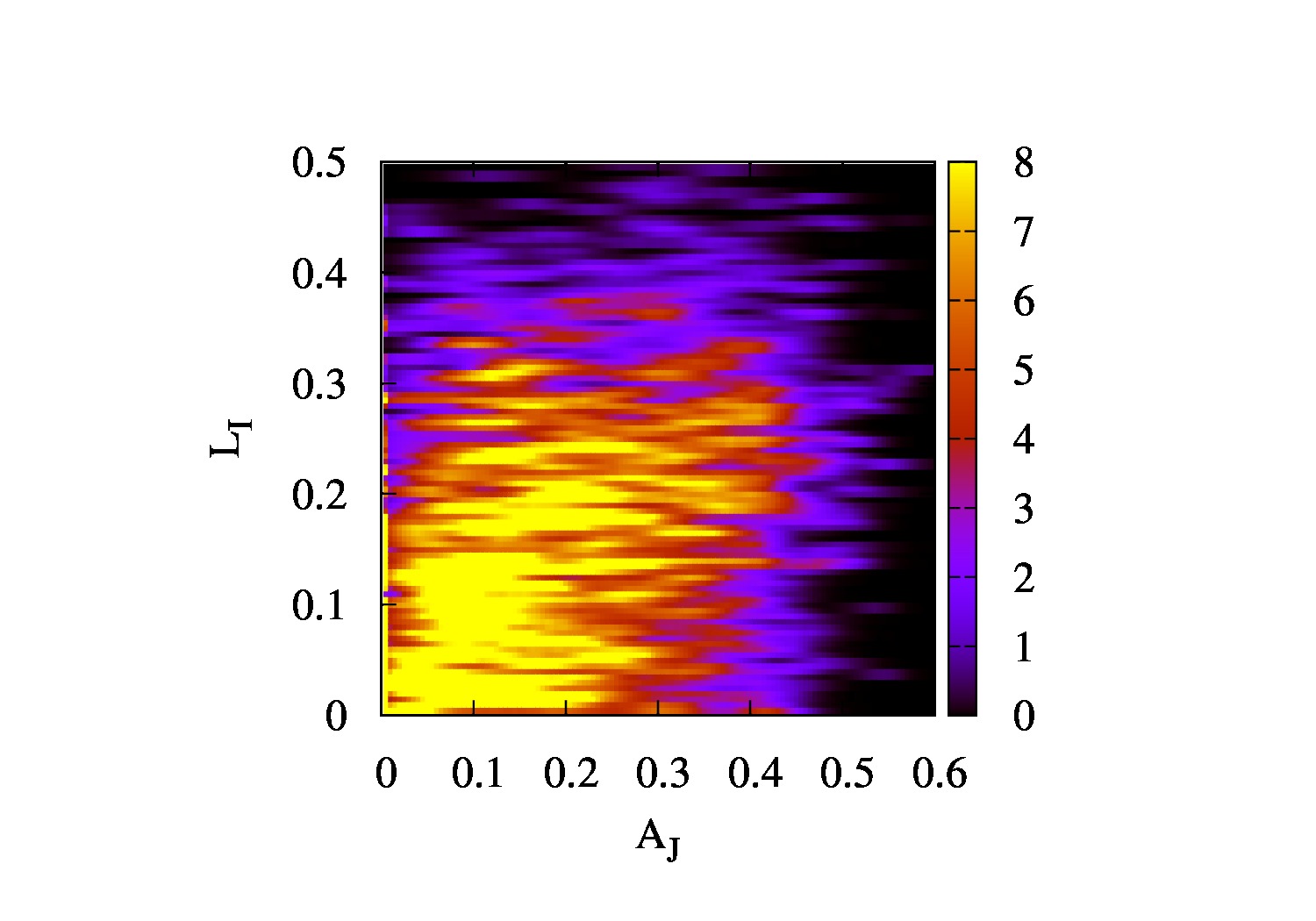}
\put(40,60){preliminary} 
\end{overpic}
\caption{Distribution of length imbalance $L_i$ depending on $A_J$.}
\label{fig:AjLi}
\end{figure}

In \ProtonProton{} collisions, where no medium creation is
expected, these splitting processes already lead to an imbalance in the
momenta of the reconstructed jets with the two highest transverse
momenta. These jets are associated with the initial back-to-back
parton pair and momentum asymmetry is caused by stochastic
distributed vacuum splitting processes out of the considered jet cone.
Experimental results \cite{ATLAS2010,CMS2011,CMS2012} in
$\sqrt{s_{NN}} = 2.76 \TeV$ \PbPb{} collisions
at the LHC showed an enhancement of this momentum imbalance in
central HIC in comparison to \ProtonProton{}-collisions.
As a measure of this enhancement the momentum imbalance $A_J$,
\begin{align}
 A_J = \frac{p_{\text{t;Leading}}-p_{\text{t;Subleading}}}{p_{\text{t;Leading}}+p_{\text{t;Subleading}}} \, ,
\end{align}
is defined, where $p_{\text{t;Leading}}$ ($p_{\text{t;Subleading}}$)
is the reconstructed transverse momentum of the jet with the highest
(second highest) transverse momentum. The additional suppression of
balanced events in HIC are supposed to be the result of different
in-medium energy loss of the two partons within the created
bulk medium, which is a consequence of a non-central spatial
production point of the initial dijet pair.

In this section we  present our first preliminary results on 
momentum imbalance simulated within the transport model BAMPS.
For the initial momentum spectra of the partons we use a distribution
sampled according to a parametrized parton distribution \cite{Glueck1995},
starting at $p_{t;0} = 100 \GeV$. Because BAMPS
describes only scattering processes of particles on the mass-shell,
it is necessary to model the initial splitting processes of the
virtual partons properly for reproducing the findings in \ProtonProton{}
collisions. Therefore the shower routines of PYTHIA \cite{Mrenna2006}
are used to model the virtual splitting processes. Because the medium
modification of the created parton showers is to be evaluated within
the BAMPS framework, it is necessary to switch off hadronization
processes and terminate the splitting processes within PYTHIA
prematurely. Therefore, the standard PYTHIA global termination
criterion in the virtuality $Q_0 = 1 \GeV$ is
replaced by an energy-dependent minimum virtuality scale
$Q_0 = \sqrt{\frac{E_{\text{parton}}}{\tau}}$ depending on the
individual parton energy and a global shower time $\tau$. Throughout
this section the shower time is assumed as $\tau = 0.2 \fm$.
Calculations within a static medium showed that the energy loss of the
reconstructed jets is, for realistic values of $\tau$, nearly independent
of the used shower time. The initial spatial production points of the
parton pairs are determined by a Glauber modelling of the initial
nucleus-nucleus collisions based on a Woods-Saxon density profile.

The so created parton showers are evolved within an offline recorded
BAMPS background event. At every timestep the shower particles can
interact with medium particles which then become shower particle
by their own. With this procedure it is possible to clearly discriminate
between shower and background particles.

In the following we compare our simulations with the
experimental data measured by CMS. All event trigger conditions by CMS
($p_{\text{t;Leading}}> 120 \GeV$, $p_{\text{t;Subleading}}>50\GeV$, $\Delta \phi > \frac{2 \pi}{3}$
and $|\eta_{\text{jet}}|<2$) and an effective handling of the detector
response and background fluctuations were used. For that an independent
Gaussian smearing of the reconstructed jet momenta is applied. The width
$\sigma$ is chosen in such a way that the smeared hadronic PYTHIA events
without shower termination can reproduce the measured \ProtonProton{}
data by CMS \cite{CMS2011}.

\Cref{fig:Aj_gauss} shows the calculated $A_J$ distribution for
central $\sqrt{s_{NN}} = 2.76 \TeV$ \PbPb{}
collisions (0-10\%, which corresponds to a mean impact parameter
$b=3.4 \fm$) with and without smearing of the reconstructed jet momenta.
As one can see, already the ``true'' jet momenta lead to an increase in the momentum asymmetry,
though it is insufficient to reproduce the measured experimental data at $A_J > 0.3$. Therefore it can be assumed that
the background fluctuations of the medium and the detector response play significant roles in explaining the strong imbalance in dijet
momenta.

One advantage of simulations within a full 3+1D transport model
is the  availability of microscopic particle informations like
space and momentum coordinates at every timestep. With this information
it is possible to further investigate the processes leading to the
observed momentum imbalance. The imbalance of the in-medium
path lengths of the parton pair is studied introducing the length
imbalance observable $L_i$
\begin{align}
 L_i = \frac{L_{\text{long}}-L_{\text{short}}}{L_{\text{long}}+L_{\text{short}}} \, .
\end{align}
Lower values of $L_i$ correspond to equal paths of the partons inside
the medium and thus more central production of the partons. The in-medium
path lengths of the initial partons ($L_{\text{long}}/L_{\text{short}}$)
are determined by their spatial production point, their initial
transverse momentum direction and their distance to the Wood-Saxon surface.
\Cref{fig:AjLi} shows the distribution of the lenght imbalance in bins of $A_J$. 
We observe that the length imbalance seems to be correlated
to the underlying momentum asymmetry. One can
state that the different transverse momenta of the reconstructed jets
are mainly caused by the different in-medium path length of the two
initial partons and hence a different energy loss. This suggests that
there are events in which the parton pairs are produced in a more
peripheral region so that one parton has to travel a longer distance
through the medium than the other one, before leaving the collision
zone.

We showed that the observation of a momentum imbalance by CMS is in
agreement with simulated BAMPS events. Therefore we showed that the consideration
of background fluctuations and detector responses plays significant role. This 
momentum imbalance is caused by a different
in-medium path length of the two initial partons. Recent experimental
results by CMS \cite{CMS2012} with a lower cone radius $R=0.3$
and lower subleading jet trigger $p_{\text{t;Subleading Jet}}>30\GeV$
show a broader and flatter $A_J$ distribution which can only be
explained within BAMPS by usage of a higher cone radius ($R=0.5$).
This implies further investigations of the influence of bulk particles
on the momentum of the reconstructed dijets. In addition, to understand
the excess on energy loss of single hadrons simulated within BAMPS, it is
highly necessary to study the relation between the momentum imbalance
$A_J$ and the nuclear modification factor $R_{AA}$.

\section{Transition from ideal to viscous Mach cones in BAMPS}
\label{section_machCones}

Highly energetic partons propagating through
the hot and dense QGP rapidly lose their energy and momentum as the energy
is deposited in the medium. Measurements of two- and three-particle
correlations in heavy-ion collisions show a complete
suppression of the away-side jet, whereas for lower $p_T$ a double
peak structure is observed  in the two-particle correlation
function \cite{Wang:2004kfa}. One possible and promising origin of these structures
is assumed to be the interaction of fast partons with the soft matter
which generates collective motion of the medium in form of Mach cones.
\cite{Stoecker:2004qu,Bouras:2010nt}.

For this purpose we investigate the propagation and formation of
Mach cones in the microscopic transport model BAMPS
(Boltzmann Approach of MultiParton Scatterings) \cite{Xu:2004mz}
in the limit of vanishing mass and very small shear viscosity over
entropy density ratio $\eta/s$ of the matter. Two different scenarios
for the jet are used. In addition, by adjusting $\eta/s$, the
influence of the viscosity on the profile of the Mach cone and
the corresponding two-particle correlation is explored for the
first time. The results presented are based on a recent publication
\cite{Bouras:2012mh}.

\begin{figure*}[ht]
\centering 
\includegraphics[width=\textwidth]{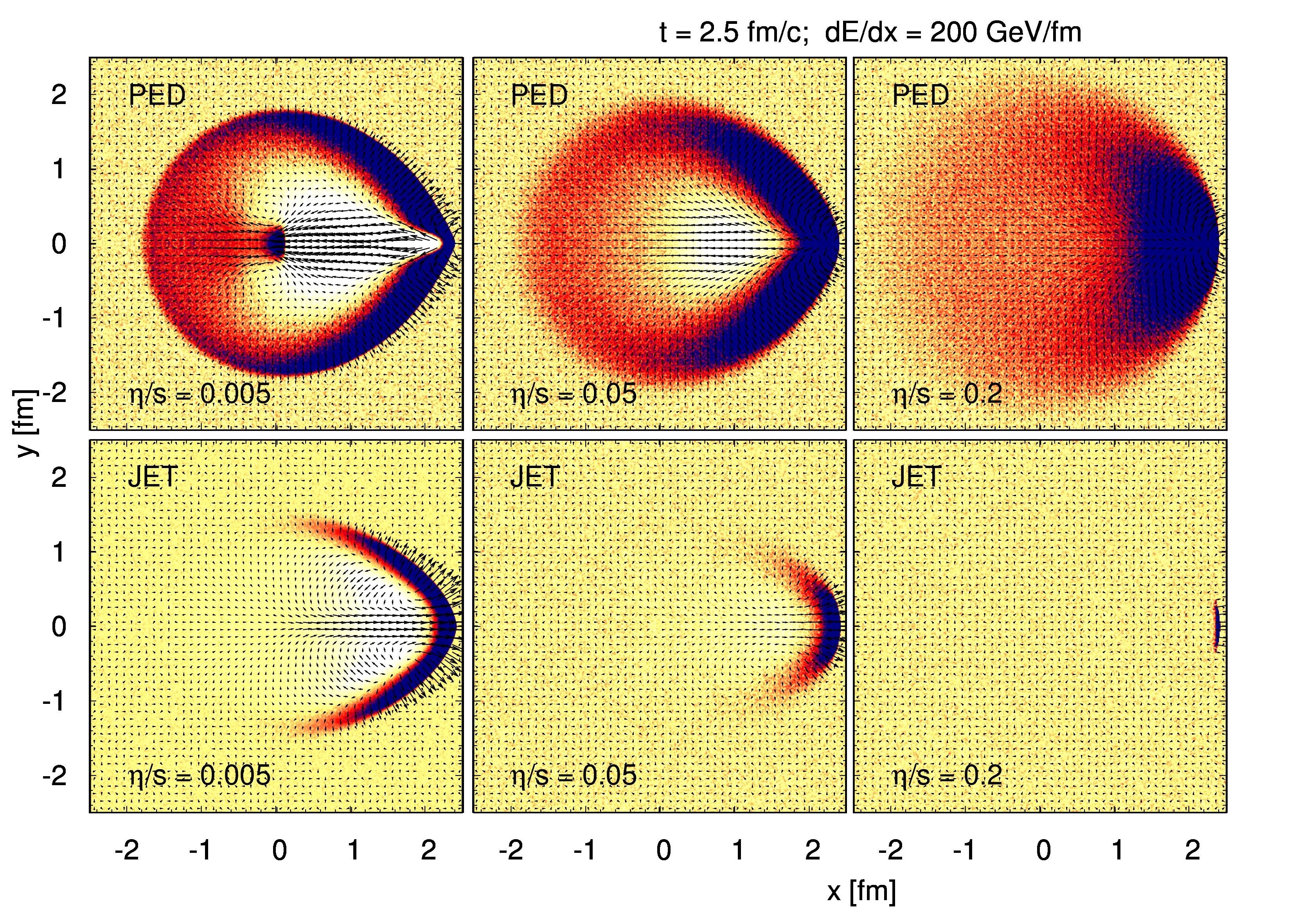}
\caption{(Color online) Transition from ideal to viscous Mach cones.
Shape of a Mach cone shown for different jet scenarios and different
viscosity over entropy density ratios, $\eta/s = 0.005$, $0.05$ and $0.5$.
The energy deposition is $dE/dx = 200$ GeV/fm. The upper
panel shows the pure energy deposition scenario (PED); the lower
panel shows the propagation of a highly energetic jet (JET) depositing
energy and momentum in $x$-direction. Depicted are the LRF energy density
within a specific range; as an overlay we show the velocity profile with
a scaled arrow length. The results are a snapshot of the
evolution at $t = 2.5$ fm/c.}
\label{fig:eDensity_viscDifSourceTerms}
\end{figure*}

Shock waves are phenomena which have their origin in
the collective motion of matter. In a simplified 
one-dimensional setup shock waves have already been studied 
within the framework of BAMPS for the perfect fluid limit
\cite{Bouras:2009nn,Bouras:2010hm}. Furthermore BAMPS
calculations have demonstrated  that the shock profile
is smeared out when viscosity is large.
It was also found that a clear observation of the shock within
the short time available in HIC requires a small viscosity.

In the following we study the evolution of "Mach cone"-like
structures with different scenarios of the jet-medium interaction
by using the parton cascade BAMPS. We focus on investigation
of Mach cone evolution in absence of any other effects
- i.e.\ we neglect such effects as initial fluctuations or
expansion, which are however relevant in HIC.
We use a static box with $T_{\rm med} = 400$ MeV and
binary collisions with an isotropic cross section.
Furthermore, we keep the mean free path $\lambda_{\rm mfp}$ of the medium
particles constant in all spatial cells by adjusting the
cross section according to $\sigma = 1 / (n\lambda_{\rm mfp})$,
where $n$ is the particle density. The related shear viscosity
for isotropic binary collisions is given by
$\eta = 0.4\,e \, \lambda_{\rm mfp}$ \cite{deGroot}.

\begin{figure}[h]
\centering 
\includegraphics[width=\columnwidth]{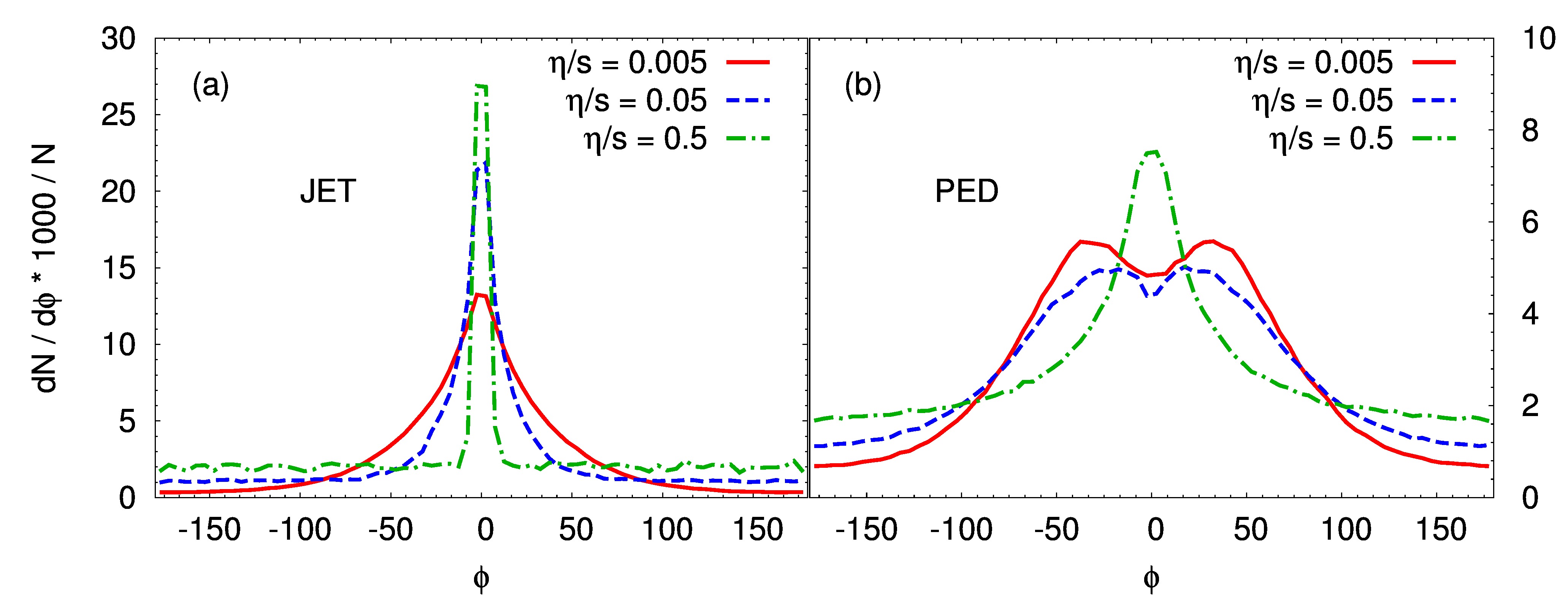}
\caption{(Color online) Two-particle correlations
$dN/(N d\phi)$ for different viscosities extracted from
calculations shown in Fig.~\ref{fig:eDensity_viscDifSourceTerms}.
The results are shown in the for the JET (a) and PED (b)
scenario for $dE/dx = 200$ GeV/fm.
}
\label{fig:num_visc_TPC}
\end{figure}

The Mach Cones studied here are induced by two different sources.
The first of them we refer to as the pure energy
deposition scenario (PED) \cite{Betz:2008ka}. This is simulated
by a moving source depositing momentum end energy isotropically
according to the thermal distribution $f(x,p) = exp(-E/T)$.
The second source we refer to as JET. This is simulated by a
highly massless particle (jet) which has only
momentum in $x$-direction, i.e. $p_{\rm x} = E_{ \rm jet}$.
After each timestep the energy of the jet particle is reset
to its initial value. For both scenarios the
sources are initialized at $t = 0$ fm/c at the position
$x = - 0.1$ fm and propagate in $x$-direction with
$v_{\rm source} = 1$, i.e. with the speed of light.

In Fig.~\ref{fig:eDensity_viscDifSourceTerms} we show the
Mach Cone structure for both PED scenario (upper panel) and
JET scenario (lower panel) with $\eta/s = 0.005$, $0.05$ and $0.5$
from left to right, respectively. We show a snapshot at $t = 2.5 \rm fm/c$.
The energy deposition rate is
fixed to $dE/dx = 200$ GeV/fm. In both scenarios, PED and JET,
for $\eta/s = 0.005$ (left panel), we observe a conical structure,
but with obvious differences. The PED case with the isotropic energy deposition
induces a spherical shock into back region; this structure is missing
in the JET scenario because of the high forward peaked momentum deposition.
Another difference is that in the JET scenario a clearly visible
head shock appears. This in turn is missing in the PED scenario.
Furthermore a (anti)-diffusion wake is induced by the JET (PED)
scenario. 

Adjusting the shear viscosity over entropy density ratio
$\eta/s = 0.05 - 0.5$ we observe a smearing out of the Mach
cone structure. For a sufficient high $\eta/s = 0.5$
the conical structure in both scenarios disappears.
This is true for shock fronts as well as for the
(anti-) diffusion wake. The difference between the PED and
the JET case is that as $\eta/s$ increases, in the PED
scenario the resulting "Mach cone" solution covers
approximately the same spatial region regardless of a value
of $\eta/s$, while in the JET case the structure is
concentrated more and more near the projectile as the
viscosity increases.

In Fig.~\ref{fig:num_visc_TPC} we show the two-particle correlations
extracted from BAMPS calculations of the Mach Cones shown in
Fig.~\ref{fig:eDensity_viscDifSourceTerms}. For the JET scenario (a)
and sufficiently small $\eta/s = 0.005$ we observe only a peak
in direction of the jet. The typical double peak structure, which
has been proposed as a possible signature of the Mach cone in HIC,
can only be observed for the PED scenario (b) and small $\eta/s$.
However, the PED scenario has no correspondence in heavy-ion physics.
We conclude that Mach cones can not be connected to double peak
structures by any realistic picture of jets in HIC. In addition, finite
values of the $\eta/s$ destroy any kind of Mach cone signatures.

\section*{Acknowledgements}

The authors are grateful to the Center for the Scientific 
Computing (CSC) at Frankfurt for the computing resources.
This work was supported by the Helmholtz International Center
for FAIR within the framework of the LOEWE program 
launched by the State of Hesse.


\end{document}